# Pressure-induced transition from $J_{\text{eff}}$=1/2 to $S$=1/2 states in CuAl$_2$O$_4$

Hwanbeom Cho[1,2,3], Choong H. Kim[2,3], Yongmoon Lee[4], Kazuki Komatsu[5], Byeong-Gwan Cho[6], Deok-Yong Cho[7], Taehun Kim[2,3,8], Chaebin Kim[2,3,8], Younghak Kim[6], Tae Yeong Koo[6], Yukio Noda[9], Hiroyuki Kagi[5], Daniel I. Khomskii[10], Donghoon Seoung[11#], and Je-Geun Park[2,3,8$]

[1]Clarendon Laboratory, University of Oxford, Parks Road, Oxford OX1 3PU, United Kingdom

[2]Center for Correlated Electron Systems, Institute for Basic Science (IBS), Seoul 08826, Republic of Korea

[3]Department of Physics and Astronomy, Seoul National University, Seoul 08826, Republic of Korea

[4]Department of Geological Sciences, Pusan National University, Busan 46241, South Korea

[5]Geochemical Research Center, Graduate School of Science, The University of Tokyo, Hongo, Bunkyo-ku, Tokyo, 113-0033, Japan

[6]Pohang Accelerator Laboratory, POSTECH, Pohang 37673, Republic of Korea

[7]IPIT & Department of Physics, Jeonbuk National University, Jeonju 54896, Republic of Korea

[8]Center for Quantum Materials, Seoul National University, Seoul 08826, Republic of Korea

[9]Institute of Multidisciplinary Research for Advanced Materials, Tohoku University, Sendai 980-8577, Japan

[10]II. Physikalisches Institut, Universität zu Köln, D-50937 Köln, Germany

[11]Department of Earth and Environmental Sciences, Chonnam National University, Gwangju 61186, South Korea

\# dseoung@jnu.ac.kr

$ jgpark10@snu.ac.kr

**ABSTRACT**

The spin-orbit entangled (SOE) $J_{\text{eff}}$-state has been a fertile ground to study novel quantum phenomena. Contrary to the conventional weakly correlated $J_{\text{eff}}$=1/2 state of 4$d$ and 5$d$ transition metal compounds, the ground state of CuAl$_2$O$_4$ hosts a $J_{\text{eff}}$=1/2 state with a strong correlation of Coulomb $U$. Here, we report that surprisingly Cu$^{2+}$ ions of CuAl$_2$O$_4$ overcome the otherwise usually strong Jahn-Teller distortion and instead stabilize the SOE state, although the cuprate has relatively small spin-orbit coupling. From the x-ray absorption spectroscopy and high-pressure x-ray diffraction studies, we obtained definite evidence of the $J_{\text{eff}}$=1/2 state with a cubic lattice at ambient pressure. We also found the pressure-induced structural transition to a compressed tetragonal lattice consisting of the spin-only $S$=1/2 state for pressure higher than $P_c$=8 GPa. This phase transition from the Mott insulating $J_{\text{eff}}$=1/2 to the $S$=1/2 states is a unique phenomenon and has not been reported before. Our study offers a rare example of the SOE $J_{\text{eff}}$-state under strong electron correlation and its pressure-induced transition to the $S$=1/2 state.

The spin-orbit coupling and a resulting quantum entangled state have been exciting research areas in condensed matter physics. Thanks to the extensive studies made over the past decade or so, it is now well known how the spin-orbit coupling governs $5d$ transition metal oxides' physical properties, particularly iridates [1-3]. Less well-known is though how this spin-orbit physics plays out under a strong electron correlation, i.e., at the regime of considerable Coulomb interaction. This question of strongly correlated spin-orbit physics seems to require two self-contradicting conditions. One is a sizeable spin-orbit coupling, which is inherently preferred for heavier elements like Ru and Ir [4]. The other is a strong electron correlation, which is favorable for lighter elements like $3d$ transition metals [5]. Hence, progress has been painfully slow in this quest despite much anticipation of rich and exotic physics.

When $d$-orbitals are put under a cubic environment, the degenerate $t_{2g}$ orbitals have the effective orbital angular momentum $L_{eff}$, leading to a spin-orbit entangled (SOE) $J_{eff}$-state [1-3] with the entangled form of the spin and the orbital wave functions. Due to the partially quenched orbital angular momentum, the SOE state system has been an excellent playground to promote exotic quantum magnetism. In particular, the SOE $J_{eff}=1/2$ state can lead to exchange anisotropy – a bond-dependent Ising-type to host the Kitaev model [6-9]. A Mott insulating phase of the $J_{eff}=1/2$ can be induced by a delicate balance among bandwidth $W$, Coulomb interaction $U$, and spin-orbit coupling (SOC) $\lambda$ [1,2]. As such, there have been reports of phase transitions induced by changing the balance via controlling the physical variables such as temperature and pressure in the systems with the SOE state [10-14].

Most studies on the spin-orbit entangled states have been so far focused on $4d$ and $5d$ transition metal compounds because the strong SOC ($\lambda$=100-400 meV) [1,2,15,16] has been widely accepted as the most critical necessary condition to create such SOE states [17]. However, it is equally interesting to ask whether SOE states can be stabilized even in $3d$ transition metal compounds. Suppose they exist, they are anticipated to be quite different from those of the conventional $J_{eff}$-states of $4d$ and $5d$ compounds with the relatively weak correlation ($U$=1.5-2 eV) [1,2,15]. In that case, the $J_{eff}$-states of $3d$ compounds will retain effects due to a strong correlation ($U$>5 eV). Until now, this is a barely explored area of the phase diagram for different ratios of $U$ and $\lambda$ [1].

Our previous experimental works [18,19] and theoretical study [20,21] suggested a cuprate system $CuAl_2O_4$ (with $Cu^{2+}$ in tetrahedra, having $e^4t_2^5$ electron configuration) as a strong candidate hosting the $J_{eff}=1/2$ state. As the $Cu^{2+}$ ion has the largest $U$ and $\lambda$, among other magnetic $3d$ transition metals [4,5], the cuprate compound would be an excellent example to introduce the SOE state with a strong correlation. In this work, using experimental techniques such as x-ray absorption spectroscopy and high-pressure x-ray diffraction, we present the definite evidence of the $J_{eff}=1/2$ state in $CuAl_2O_4$ at ambient pressure and a structural transition to the state with $S$=1/2 at high pressure.

$CuAl_2O_4$ has a spinel structure (Fig. 1a), where $Cu^{2+}$ is located at the center of the tetrahedron ($A$-site) [18,19]. $Al^{3+}$ sits at the center of the octahedra that consists of six neighboring oxygen ions ($B$-site). Due to the $T_d$ point group symmetry of the crystal field, the 3$d$ orbitals of $Cu^{2+}$ ion in the tetrahedral site are split into upper $t_2$ and lower $e$ orbitals. As depicted in the left-hand-side of Fig. 1b, one hole in the $t_2$ levels has $l_{eff}$=1. Note that capital $J_{eff}$, $L_{eff}$, and $S$ ($j_{eff}$, $l_{eff}$, and $s$) stand for multi-particle (single-particle) total, orbital, and spin angular momenta. The SOC further splits the $t_2$ levels into the SOE $j_{eff}$=1/2 doublet occupied by the single hole and the $j_{eff}$=3/2 states fully occupied by electrons. Finally, Coulomb interactions can give rise to a Mott insulating $J_{eff}$=1/2 state by splitting the $j_{eff}$=1/2 band.

However, on the other limit (right-hand-side of Fig. 1b), where the energy scale of Jahn-Teller distortion $\Delta$ is larger than $\lambda$, an $S$=1/2 state with fully quenched orbital angular momentum will be instead stabilized. At ambient pressure, our DFT and DMFT calculations [20] predicted that in $CuAl_2O_4$, the $J_{eff}$=1/2 state wins over the spin-only state. Although SOC is relatively small ($\lambda$=50 meV) for $Cu^{2+}$ ions, the strong correlation ($U$=7eV) works as if SOC is effectively enhanced to stabilize the $J_{eff}$=1/2 state. Since the $j_{eff}$=1/2 state occupied by the single hole involves the equal amount of $t_2$ orbitals ($\alpha'$=1/3)

$$\left|j_{eff} = \tfrac{1}{2}, j^z_{eff} = \pm\tfrac{1}{2}\right\rangle = \sqrt{\alpha'}|l^z_{eff} = 0\rangle|\pm\rangle - \sqrt{1-\alpha'}|l^z_{eff} = \pm 1\rangle|\mp\rangle \quad (1)$$

, where $|l^z_{eff} = 0\rangle = |d_{xy}\rangle$, $|l^z_{eff} = \pm 1\rangle = -\tfrac{1}{\sqrt{2}}(i|d_{zx}\rangle \pm |d_{yz}\rangle)$ and $|\pm\rangle$ is the spin-half spinor [6], the $CuO_4$ tetrahedron is not distorted into $D_{2d}$ symmetry but sustains instead the cubic ($T_d$) symmetry. The cubic symmetry of $CuAl_2O_4$ at ambient pressure was verified via low-temperature neutron- [18] and x-ray diffraction [19] methods. But there is still a question of locally broken symmetry, which motivated this work together with a possible pressure-induced new transition.

XANES measurement was done at the copper K-edge to estimate the accurate value for site-disorder in the single-crystal. The other purpose was to confirm the lattice symmetry, i.e., the cubic structure without Jahn-Teller distortion. The observed spectrum in Fig. 2a has ten peaks, which we can assign as peaks A to J. We can simulate the total spectrum by superposing two spectra, calculated with $Cu^{2+}$ in the tetrahedral and octahedral sites of the cubic lattice, respectively, in the ratio of 64:36. The simulation's overall shape consists of ten peaks corresponding to the measurement, which indicates that the crystal indeed has the cubic lattice with the site-disorder of $\eta$=0.36, comparable to that obtained by the Rietveld method of the previous x-ray diffraction data [19].

The tetrahedral site with $T_d$ symmetry can form a $p$-$d$ hybridized orbital, enabling the electric dipolar transition of 1$s$ electron to the hybridized orbital [22,23]. The pre-edge peak A comes from a transition to the copper 3$d$-4$p$ hybridized orbital in the tetrahedral site. In contrast, peak B originates from the hybridization of copper 4$p$, oxygen 2$p$, and aluminum 3$s$ orbitals. On the other hand, the peaks C, D, F, H, J (E, G, I) arise from transitions to the copper $p$ orbital states in the tetrahedral (octahedral) sites. This result is similar to the study on $CuAl_2O_4$ nano-particles [24].

The intensity of the peaks in the experimental spectrum looks suppressed when compared with that in the calculation. This is due to the self-absorption effect [25], which occurs when a single-crystal is measured in a fluorescence yield mode. Moreover, since the simulation is based on a Hartree-Fock calculation with a muffin-tin potential [26], electron correlation is difficult to estimate precisely. This limitation can explain a difference in the position of the 3$d$ band.

Copper L-edge XAS probes more directly the spin-orbit entangled character of the unoccupied state in CuAl$_2$O$_4$. This technique has so far been proven a useful tool to investigate the SOC of a ground state by using the electric dipole selection rules [15,27-29]. The $j_{eff}$=1/2 state is branched off from the atomic $j$=5/2, $|j_{eff} = 1/2, j^z_{eff} = \pm 1/2\rangle \propto \pm 1/\sqrt{6}|j = 5/2, j_z = \pm 5/2\rangle \mp 5/\sqrt{30}|j = 5/2, j_z = \mp 3/2\rangle$. If the hole in CuAl$_2$O$_4$ occupies the $j_{eff}$=1/2 state, the transition from the 2$p_{1/2}$ to $j_{eff}$=1/2 states (L$_2$-edge) is forbidden while the transition from the 2$p_{3/2}$ to $j_{eff}$=1/2 states (L$_3$-edge) is allowed. On the other hand, the transitions at both edges are allowed for the hole occupying a spin-1/2 state with fully quenched orbital angular momentum. Therefore, if the ground state of CuAl$_2$O$_4$ is indeed $J_{eff}$=1/2, we expect to observe a single peak at the L$_3$-edge and no peak at the L$_2$-edge in the absorption spectrum.

However, there is a sizable amount of the site-disorder in our single-crystal sample: 64 (36) % of Cu$^{2+}$ locate in the tetrahedral (octahedral) site. Contrary to the octahedral site, where the hole occupies the $e_g$ orbital state with $L_{eff}$=0, the tetrahedral site can solely host the $J_{eff}$=1/2 state. This will make an absorption peak $\alpha$ at the L$_3$-edge in the tetrahedral site, but the octahedral site will make two absorption peaks $\beta$ and $\gamma$ at the L$_3$- and L$_2$-edges, respectively (Fig. 2b). Therefore, we expect to observe a three-peaks feature, two peaks at the L$_3$-edge and a single peak at the L$_2$-edge, in the absorption spectrum.

From the XAS measurement of a single-crystal CuAl$_2$O$_4$, we verified that there are indeed two distinct peaks at the L$_3$-edge (Fig. 2c) and a single peak at the L$_2$-edge (Fig. 2d) as expected. At the L$_3$-edge, the ratio of the spectral weight of peaks $\alpha$ and $\beta$ is estimated to be 0.636:0.364 [30], which is comparable to the above result of $\eta$=0.36 obtained from the XANES experiment. The spacing between the two peaks, 0.6(1) eV corresponds well to the energy difference $\Delta_\epsilon$=0.7 eV between the first conduction band projected onto the $j_{eff}$=1/2 state in the tetrahedral site and the second band projected onto the $e_g$ state in the octahedral site [30]. We can exclude the presence of Cu metal, Cu$^+$ and Cu$^{3+}$ ions, from the possible reason for the peak splitting because of the observed small splitting energy: for the latter cases, one expects to see much larger splitting in the order of a few eV [31,32].

Moreover, the transition ratio at the L$_3$- and L$_2$-edges, or the branching ratio (BR), gives the expectation value <$L \cdot S$> of the ground state and <$L \cdot S$> becomes zero for BR=2 [28]. Therefore, in the octahedral site, the ratio of the spectral weight of peaks $\beta$ and $\gamma$ should follow the ratio of 2:1. The peak $\gamma$ in Fig. 2d calculated using the half spectral weight of peak $\beta$ is indeed comparable to the observed spectrum [30]. Finally, by subtracting the octahedral site's contribution (peak $\beta$ and $\gamma$) from the observed spectrum, we obtained the contribution solely from the tetrahedral site [30]. Using the resultant spectral weight at each edge, we evaluated BR in the tetrahedral site. It gives <$L_{eff} \cdot S$>=0.90(8), which is close to <$L_{eff} \cdot S$>=1 of the $J_{eff}$=1/2 state [28,33].

The spin-orbit entangled $J_{eff}$=1/2 state of CuAl$_2$O$_4$ arises from the delicate balance among $W$, $U$, and $\lambda$ [1,2]. Therefore, one can expect to induce various transitions of the SOE states according to the values of $W$, $U$, and $\lambda$ by modifying the balance via adjusting a physical variable such as pressure[10-14]. Since the $J_{eff}=1/2$ state of $CuAl_2O_4$ is in the exotic region of the phase diagram of $U$ and $\lambda$, where large $U$ and small $\lambda$ make the strongly correlated spin-orbit regime [30], a novel phase transition distinct from those in the other $J_{eff}=1/2$ states is expected to emerge by applying pressure.

To examine this possibility of a pressure-induced new phase, we carried out high-pressure x-ray diffraction experiments to find the pressure-induced structural transition. Fig. 3a shows that the Bragg peaks get split as pressure increases. The split Bragg peaks merge again and return to its initial phase as the pressure is released, ensuring that this pressure-induced transition is reversible. We also confirmed that the cubic lattice changes into a tetragonal lattice [30] with a space group of $I4_1/amd$ and a reduced lattice parameter $a_t=a_c/\sqrt{2}$, where $a_c$ is the lattice parameter of the cubic phase.

According to our experimental data and subsequent analysis, $a_c$ gets continuously reduced as pressure increases to the critical pressure $P_c\sim 8$ GPa (Fig. 3b). The pressure dependence of $a_c$ follows the typical Vinet equation of states [34] with a bulk modulus of $B_0=247(7)$ GPa and $dB_0/dP=7$, comparable to those of other spinel compounds [35]. Above $P_c$, the structural transition emerges and $\sqrt{2}a_t$ bifurcates from the lattice parameter along the c-axis, $c_t$. Comparing with $a_t$, which is increasing slightly as the pressure increases ($\Delta(\sqrt{2}a_t)=0.05(1)$ Å), notably $c_t$ reduces dramatically ($\Delta(c_t)=-0.28(4)$ Å). By fitting the pressure dependence of the tetragonal phase's lattice parameters using a mean-field equation, we can evaluate the critical pressure of $P_c=8(1)$ GPa.

Fig. 3c shows the pressure dependence of a ratio of the lattice parameters $c/a$, where we plot $c_t/(\sqrt{2}a_t)$ for the tetragonal phase. Above $P_c$, the ratio drops down to 0.979(3) and keeps reducing while the pressure increases. The reduction of $c/a$ comes mainly from the contraction of the $CuO_4$ tetrahedron. Depicted in Fig. 3d, O-Cu-O angle, $\phi_t$, of the tetrahedron increases from 109.5° above $P_c$. Moreover, a ratio of the height $l_c$ to the base $l_a$ of the tetrahedron follows almost the same pressure dependence of global $c/a$.

To understand the mechanism underlying the pressure-induced structural phase transition in $CuAl_2O_4$, we performed DFT calculations. The ground state of $CuAl_2O_4$ at the ambient pressure is confirmed to be $J_{eff}=1/2$ with the cubic lattice and it remains robust even with a notable amount of site-disorder $\eta=0.5$ [30]. The total energy calculated from the $CuAl_2O_4$ lattice shows its global minimum locating at the point where the ratio of the lattice parameters $c/a$ is unity (Fig. 4a). However, as the pressure increases, the unit cell volume reduces, and the bond length of copper and oxygen ions becomes shorter. The lattice optimization calculations show the shift of the balance from the undistorted $J_{eff}=1/2$ state to the state with $S=1/2$ with Jahn-Teller distortion. As a result, the minimum of the total energy shifts suddenly towards $c/a<1$ above the theoretical critical volume change of the unit cell $1-(V_c/V_0)=0.02$.

In Fig. 4b, the mixing parameter and the lattice parameters' ratio reach a value of $\alpha'=1/3$

and $c/a$=1, respectively below the volume change of 2 %, which denotes the stabilization of the $J_{eff}$=1/2 state with the cubic lattice. When the volume changes more than 2 %, however, $α'$ moves towards 1 and $c/a$<0.95, suggesting that the disentangled $S$=1/2 state with the compressed tetragonal lattice has been stabilized for the phase with a smaller volume. This kind of a pressure-dependent quantum phase transition has not been previously reported in other systems with $J_{eff}$=1/2 state [10-14].

The cubic lattice's observed transition into the compressed tetragonal lattice corresponds to what we expected from the total energy calculation (Fig. 4c). Therefore, our measurement supports the scenario, where the ground state of $CuAl_2O_4$ changes from the SOE $J_{eff}$=1/2 to the spin-only $S$=1/2 state. The behavior of $c/a$ depending on the volume change corresponds to the theoretical expectation within the error bar. There is a slight difference in the estimate of the critical value $1-(V_c/V_0)$ between the experimental (2.9(6) %) and the theoretical (2 %) values. It can be due to the limitation or error in determining $U$, which directly relates to the values of $V_c$ and $c/a$ in our method.

From the x-ray absorption and high-pressure XRD studies, we observed the direct evidence of the $J_{eff}$=1/2 state in $CuAl_2O_4$ at ambient pressure, leading to the cubic lattice. And we also discovered the pressure-induced structural transition to the compressed tetragonal lattice, hosting the spin-only state. This observation corresponds well to what we expect from the $J_{eff}$=1/2 state under the regime of strong correlation. The crystal structure of $CuAl_2O_4$, where the separation of tetrahedra and the large $U$ cooperate in stabilizing the $J_{eff}$=1/2 state of $Cu^{2+}$, also allows the pressure-dependent evolution of the $J_{eff}$=1/2 state in an isolated "cage." Our finding of the strongly correlated SOE state in $CuAl_2O_4$ can thus provide insight into unexplored territory in the phase diagram of $U$ and $λ$, where $3d$ transition metal ions stabilize the SOE state by competing with Jahn-Teller distortion. Another candidate with a similar effect may include $Co^{2+}$ in an octahedral crystal field, which can host the Kitaev model with the $J_{eff}$=1/2 state [36,37]. For instance, the $CoO_6$ octahedra of $γ$-$SiCo_2O_4$ [38,39] and $Ba_3CoSb_2O_9$ [40,41] are reported to have a $D_{3d}$ point group symmetry instead of a $C_{2h}$ driven by Jahn-Teller distortion. It is noted that another spinel compound $NiRh_2O_4$ [42], where $Ni^{2+}$ is positioned at the tetrahedral site, yields magnetic entropy of $R\ln6$ instead of $R\ln3$ of spin only $S$=1, which implies a ground state with entanglement.

It should be noted that the transition of the ground state from $J_{eff}$=1/2 to $S$=1/2 has not been reported in any other systems before. For instance, in the Ruddlesden-Popper series of iridates, $Sr_{n+1}Ir_nO_{3n+1}$ [10,11], the pressure broadens the bandwidth of $t_{2g}$ states and eventually turns the ground state into a typical metal. In the case of the honeycomb-based Kitaev candidates, $α$-$Li_2IrO_3$ [12], $β$-$Li_2IrO_3$ [13], and $α$-$RuCl_3$ [14], the pressure leads to dimerization, changing the Mott insulating $J_{eff}$=1/2 phase to a gapped dimerized one.

Two mechanisms could, in principle, suppress the effect of SOC under pressure. One is the splitting of triply-degenerate $t_2$ levels by extra distortions, in particular, due to the Jahn-Teller effect. The other more substantial effect is the inter-site electron hopping $t$ and the corresponding formation of bands [43-45]. Since both of these effects become more critical with increasing electron hopping or the bandwidth, they will be enhanced by pressure, which could explain the novel transition from the SOE $J_{\text{eff}}$=1/2 state at ambient pressure to the Jahn-Teller distorted $S$=1/2 state at high pressure that we discovered in $CuAl_2O_4$.

In conclusion, we have successfully demonstrated that the $CuAl_2O_4$ has a $J_{\text{eff}}$=1/2 state at ambient pressure, which undergoes a rare transition to an $S$=1/2 state upon applying pressure. This pressure-driven transition between a $J_{\text{eff}}$=1/2 state and an $S$=1/2 state is well captured by the DFT calculations with the spin-orbit coupling fully employed. Our observation constitutes the first observation of pressure-driven $J_{\text{eff}}$-to-$S$ states with the intrinsically strong Coulomb interaction under strong correlation. Our work offers an exciting opportunity to explore the spin-orbit entanglement at the regime of strong correlation.

We acknowledge helpful discussions with Wondong Kim, Roger Johnson, and Frank de Groot. This work was supported by the Institute for Basic Science (IBS) in Korea [No. IBS-R009-G1 (H.C., T.K., C.K., J.G.P.) and No. IBSR009-D1 (C.H.K.)] and the Leading Researchers Program of the National Research Foundation of Korea [NRF-2020R1A3B2079375 (T.K., C.K., J.G.P.) and NRF-2019K1A3A7A09101574 (D.S.)]. H.C. acknowledges funding from the European Research Council (ERC) under the European Union's Horizon's 2020 research and innovation program Grant Agreement Number 788814, the work of D.I.Kh. was funded by the Deutsche Forschungsgemeinschaft (DFG, German Research Foundation) - Project number 277146847 - CRC 1238. The synchrotron-based experiments were done in PAL and Photon Factory (Proposal No. 2017G644).

# REFERENCES

[1] W. Witczak-Krempa, G. Chen, Y. B. Kim, and L. Balents, Correlated quantum phenomena in the strong spin-orbit regime, Annu. Rev. Condens. Matter Phys. **5**, 57 (2014).

[2] B. J. Kim, H. Jin, S. J. Moon, J.-Y. Kim, B.-G. Park, C. S. Leem, J. Yu, T. W. Noh, C. Kim, S.-J. Oh, J.-H. Park, V. Durairaj, G. Cao, and E. Rotenberg, Novel $J_{\text{eff}}=1/2$ Mott state induced by relativistic spin-orbit coupling in $Sr_2IrO_4$, Phys. Rev. Lett. **101**, 076402 (2008).

[3] B. J. Kim, H. Ohsumi, T. Komesu, S. Sakai, T. Morita, H. Takagi, and T. Arima, Phase-sensitive observation of a spin-orbital Mott state in $Sr_2IrO_4$, Science **323**, 1329 (2009).

[4] D. I. Khomskii, *Transition Metal Compounds* (Cambridge Univ. Press, Cambridge, 2014).

[5] A. Georges, L. de Medici, and J. Mravlje, Strong correlations from Hund's coupling, Annu. Rev. Condens. Matter Phys. **4**, 137 (2013).

[6] G. Jackeli and G. Khaliullin, Mott insulators in the strong spin-orbit coupling limit: from Heisenberg to a quantum compass and Kitaev models, Phys. Rev. Lett. **102**, 017205 (2009).

[7] K. Kitagawa, T. Takayama, Y. Matsumoto, A. Kato, R. Takano, Y. Kishimoto, S. Bette, R. Dinnebier, G. Jackeli, and H. Takagi, A spin–orbital-entangled quantum liquid on a honeycomb lattice, Nature **554**, 341 (2018).

[8] Y. Kasahara, T. Ohnishi, Y. Mizukami, O. Tanaka, Sixiao Ma, K. Sugii, N. Kurita, H. Tanaka, J. Nasu, Y. Motome, T. Shibauchi, and Y. Matsuda, Majorana quantization and half-integer thermal quantum Hall effect in a Kitaev spin liquid, Nature **559**, 227 (2018).

[9] S.-H. Do, S.-Y. Park, J. Yoshitake, J. Nasu, Y. Motome, Y. S. Kwon, D. T. Adroja, D. J. Voneshen, K. Kim, T.-H. Jang, J.-H. Park, K.-Y. Choi, and S. Ji, Majorana fermions in the Kitaev quantum spin system $\alpha$-$RuCl_3$, Nat. Phys. **13**, 1079 (2017).

[10] C. Donnerer, Z. Feng, J. G. Vale, S. N. Andreev, I. V. Solovyev, E. C. Hunter, M. Hanfland, R. S. Perry, H. M. Rønnow, M. I. McMahon, V. V. Mazurenko, and D. F. McMorrow, Pressure dependence of the structure and electronic properties of $Sr_3Ir_2O_7$, Phys. Rev. B **93**, 174118 (2016).

[11] D. Haskel, G. Fabbris, Mikhail Zhernenkov, P. P. Kong, C. Q. Jin, G. Cao, and M. van Veenendaal, Pressure tuning of the spin-orbit coupled ground state in $Sr_2IrO_4$, Phys. Rev. Lett. **109**, 027204 (2012).

[12] V. Hermann, M. Altmeyer, J. Ebad-Allah, F. Freund, A. Jesche, A. A. Tsirlin, M. Hanfland, P. Gegenwart, I. I. Mazin, D. I. Khomskii, R. Valentí, and C. A. Kuntscher, Competition between spin-orbit coupling, magnetism, and dimerization in the honeycomb iridates: $\alpha$-$Li_2IrO_3$ under pressure, Phys. Rev. B **97**, 020104(R) (2018).

[13] T. Takayama, A. Krajewska, A. S. Gibbs, A. N. Yaresko, H. Ishii, H. Yamaoka, K. Ishii, N.

Hiraoka, N. P. Funnell, C. L. Bull, and H. Takagi, Pressure-induced collapse of the spin-orbital Mott state in the hyperhoneycomb iridate $\beta$-Li$_2$IrO$_3$, Phys. Rev. B **99**, 125127 (2019).

[14] G. Bastien, G. Garbarino, R. Yadav, F. J. Martinez-Casado, R. Beltrán Rodríguez, Q. Stahl, M. Kusch, S. P. Limandri, R. Ray, P. Lampen-Kelley, D. G. Mandrus, S. E. Nagler, M. Roslova, A. Isaeva, T. Doert, L. Hozoi, A. U. B. Wolter, B. Büchner, J. Geck, and J. van den Brink, Pressure-induced dimerization and valence bond crystal formation in the Kitaev-Heisenberg magnet $\alpha$-RuCl$_3$, Phys. Rev. B **97**, 241108(R) (2018).

[15] K. W. Plumb, J. P. Clancy, L. J. Sandilands, V. Vijay Shankar, Y. F. Hu, K. S. Burch, H.-Y. Kee, and Y.-J. Kim, $\alpha$-RuCl$_3$: a spin-orbit assisted Mott insulator on a honeycomb lattice, Phys. Rev. B **90**, 041112(R) (2014).

[16] A. Banerjee, C. A. Bridges, J.-Q. Yan, A. A. Aczel, L. Li, M. B. Stone, G. E. Granroth, M. D. Lumsden, Y. Yiu, J. Knolle, S. Bhattacharjee, D. L. Kovrizhin, R. Moessner, D. A. Tennant, D. G. Mandrus, and S. E. Nagler, Proximate Kitaev quantum spin liquid behaviour in a honeycomb magnet, Nat. Mater. **15**, 733 (2016).

[17] G. Cao and P. Schlottmann, The challenge of spin–orbit-tuned ground states in iridates: a key issues review, Rep. Prog. Phys. **81**, 042502 (2018).

[18] R. Nirmala, K.-H. Jang, H. Sim, H. Cho, J. Lee, N.-G. Yang, S. Lee, R. M. Ibberson, K. Kakurai, M. Matsuda, S.-W. Cheong, V. V. Gapontsev, S. V. Streltsov, and J.-G. Park, Spin glass behavior in frustrated quantum spin system CuAl$_2$O$_4$ with a possible orbital liquid state, J. Phys.: Condens. Matter **29**, 13LT01 (2017).

[19] H. Cho, R. Nirmala, J. Jeong, P. J. Baker, H. Takeda, N. Mera, S. J. Blundell, M. Takigawa, D. T. Adroja, and J.-G. Park, Dynamic spin fluctuations in the frustrated $A$-site spinel CuAl$_2$O$_4$, Phys. Rev. B **102**, 014439 (2020).

[20] C. H. Kim, S. Baidya, H. Cho, V. V. Gapontsev, S. V. Streltsov, D. I. Khomskii, J.-G. Park, A. Go, and H. Jin, Theoretical evidence of spin-orbital-entangled $J_{\text{eff}}$=1/2 state in the $3d$ transition metal oxide CuAl$_2$O$_4$, Phys. Rev. B **100**, 161104(R) (2019).

[21] S. A. Nikolaev, I. V. Solovyev, A. N. Ignatenko, V. Y. Irkhin, and S. V. Streltsov, Realization of the anisotropic compass model on the diamond lattice of Cu$^{2+}$ in CuAl$_2$O$_4$, Phys. Rev. B **98**, 201106(R) (2018).

[22] T. Yamamoto, Assignment of pre-edge peaks in K-edge x-ray absorption spectra of $3d$ transition metal compounds: electric dipole or quadrupole, X-Ray Spectrom. **37**, 572 (2008).

[23] S. E. Shadle, J. E. Penner-Hahn, H. J. Schugar, B. Hedman, K. O. Hodgson, and E. I. Solomon, X-ray absorption spectroscopic studies of the blue copper site: metal and ligand K-edge studies to probe the origin of the EPR hyperfine splitting in plastocyanin, J. Am. Chem. Soc. **115**, 767 (1993).

[24] T. Tangcharoen, W. Klysubun, and C. Kongmark, Synchrotron X-ray absorption

spectroscopy and cation distribution studies of $NiAl_2O_4$, $CuAl_2O_4$, and $ZnAl_2O_4$ nanoparticles synthesized by sol-gel auto combustion method, J. Mol. Struct. **1182**, 219 (2019).

[25] J. A. van Bokhoven and C. Lamberti, *X-Ray Absorption and X-Ray Emission Spectroscopy: Theory and Applications* (Wiley, New York, 2016).

[26] A. L. Ankudinov, B. Ravel, J. J. Rehr, and S. D. Conradson, Real-space multiple-scattering calculation and interpretation of x-ray-absorption near-edge structure, Phys. Rev. B **58**, 7565 (1998).

[27] J. P. Clancy, N. Chen, C. Y. Kim, W. F. Chen, K. W. Plumb, B. C. Jeon, T. W. Noh, and Y.-J. Kim, Spin-orbit coupling in iridium-based $5d$ compounds probed by x-ray absorption spectroscopy, Phys. Rev. B **86**, 195131 (2012).

[28] B. T. Thole and G. van der Laan, Linear relation between x-ray absorption branching ratio and valence-band spin-orbit expectation value, Phys. Rev. A **38**, 1943 (1988).

[29] D.-Y. Cho, J. Park, J. Yu, and J.-G. Park, X-ray absorption spectroscopy studies of spin–orbit coupling in 5d transition metal oxides, J. Phys.: Condens. Matter **24**, 055503 (2012).

[30] See Supplemental Material at http://link.aps.org/ for further details of the experimental and theoretical methods, and supporting data.

[31] P. Jiang, D. Prendergast, F. Borondics, S. Porsgaard, L. Giovanetti, E. Pach, J. Newberg, H. Bluhm, F. Besenbacher, and M. Salmeron, Experimental and theoretical investigation of the electronic structure of $Cu_2O$ and $CuO$ thin films on Cu(110) using x-ray photoelectron and absorption spectroscopy, J. Chem. Phys. **138**, 024704 (2013).

[32] R. Sarangi, N. Aboelella, K. Fujisawa, W. B. Tolman, B. Hedman, K. O. Hodgson, and E. I. Solomon, X-ray absorption edge spectroscopy and computational studies on $LCuO_2$ species: superoxide-$Cu^{II}$ versus peroxide-$Cu^{III}$ bonding, J. Am. Chem. Soc. **128**, 8286 (2006).

[33] J.-H. Sim, H. Yoon, S. H. Park, and M. J. Han, Calculating branching ratio and spin-orbit coupling from first principles: a formalism and its application to iridates, Phys. Rev. B **94**, 115149 (2016).

[34] P. Vinet, J. R. Smith, J. Ferrante, and J. H. Rose, Temperature effects on the universal equation of state of solids, Phys. Rev. B **35**, 1945 (1987).

[35] F. J. Manjon, I. Tiginyanu, and V. Ursaki, *Pressure-Induced Phase Transitions in $AB_2X_4$ Chalcogenide Compounds Ch. 2* (Springer, Berlin, 2014).

[36] H. Liu and G. Khaliullin, Pseudospin exchange interactions in $d^7$ cobalt compounds: possible realization of the Kitaev model, Phys. Rev. B **97**, 014407 (2018).

[37] H. Liu, J. Chaloupka, and G. Khaliullin, Kitaev Spin Liquid in $3d$ Transition Metal Compounds, Phys. Rev. Lett. **125**, 047201 (2020).

[38] F. Marumo, M. Isobe, and S. Akimoto, Electron-density distributions in crystals of $\gamma$-$Fe_2SiO_4$ and $\gamma$-$Co_2SiO_4$, Acta Cryst. B**33**, 713 (1977).

[39] W. Zhang, S. Okubo, H. Ohta, T. Saito, and M. Takano, High-frequency ESR measurements of the Co spinel compound $SiCo_2O_4$, J. Phys.: Condens. Matter **19**, 145264 (2007).

[40] Y. Doi, Y. Hinatsu, and K. Ohoyama, Structural and magnetic properties of pseudo-two-dimensional triangular antiferromagnets $Ba_3MSb_2O_9$ (M=Mn, Co, and Ni), J. Phys.: Condens. Matter **16**, 8923 (2004).

[41] Y. Shirata, H. Tanaka, A. Matsuo, and K. Kindo, Experimental realization of a spin-1/2 triangular-lattice Heisenberg antiferromagnet, Phys. Rev. Lett. **108**, 057205 (2012).

[42] J. R. Chamorro, L. Ge, J. Flynn, M. A. Subramanian, M. Mourigal, and T. M. McQueen, Frustrated spin one on a diamond lattice in $NiRh_2O_4$, Phys. Rev. Materials **2**, 034404 (2018).

[43] S. V. Streltsov and D. I. Khomskii, Covalent bonds against magnetism in transition metal compounds, Proc. Natl. Acad. Sci. USA **113**, 10491 (2016).

[44] M. Ye, H.-S. Kim, J.-W. Kim, C.-J. Won, K. Haule, D. Vanderbilt, S.-W. Cheong, and G. Blumberg, Covalency-driven collapse of strong spin-orbit coupling in face-sharing iridium octahedra, Phys. Rev. B **98**, 201105(R) (2018).

[45] D. I. Khomskii and S. V. Streltsov, Orbital effects in solids: basics and novel development, Preprint at https://arxiv.org/abs/2006.05920 (2020).

**FIGURES**

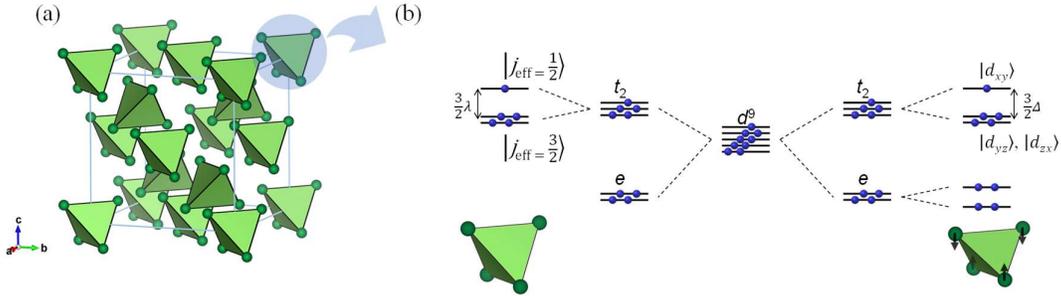

FIG. 1. (a) The unit cell structure of $CuAl_2O_4$. $Cu^{2+}$ ions are located at the tetrahedra's center made of four oxygen ions (green balls). The $Al^{3+}$ ions are omitted for simplicity. The neighboring tetrahedra do not share a common oxygen ion and are separated from each other. (b) The ground state of $CuAl_2O_4$ at different regimes. (Left-hand-side) If $\lambda > \Delta$, $t_2$ levels are split into the $j_{\text{eff}}=1/2$ and $j_{\text{eff}}=3/2$ states, and the $CuO_4$ tetrahedron retains its cubic symmetry. (Right-hand-side) If $\lambda < \Delta$, $t_2$ levels are split into states with fully quenched orbital angular momentum, and the tetrahedron becomes tetragonally compressed along its local $c$-axis.

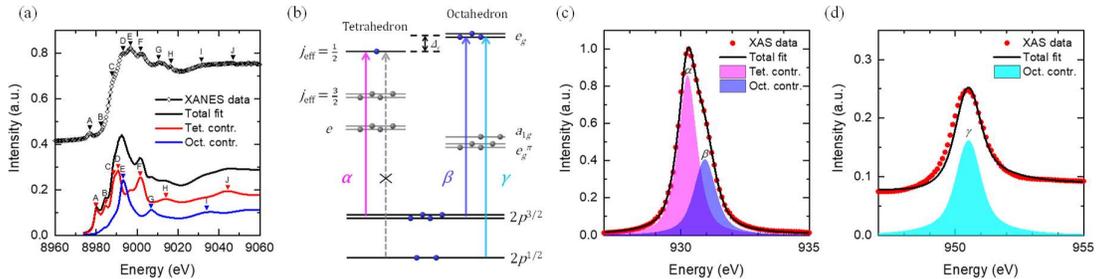

FIG. 2. (a) XANES spectrum at the copper K-edge. The total spectral calculation done using the cubic lattice is composed of two contributions: 64 (36) % of $Cu^{2+}$ ions located in the tetrahedral (octahedral) site. (b) Electric dipole transition occurring in the tetrahedral and octahedral sites. In the tetrahedral site, the transition at $L_3$-edge ($\alpha$) is allowed, but the transitions at $L_2$-edge is forbidden. However, in the octahedral site, the transitions at both $L_3$- and $L_2$-edges ($\beta$ and $\gamma$, respectively) are allowed. (c) The observed absorption spectrum at the copper $L_3$-edge (~930 eV). It has a distinct feature of two peaks: $\alpha$ and $\beta$. (d) The spectrum at the copper $L_2$-edge (~950 eV). It features a single peak $\gamma$.

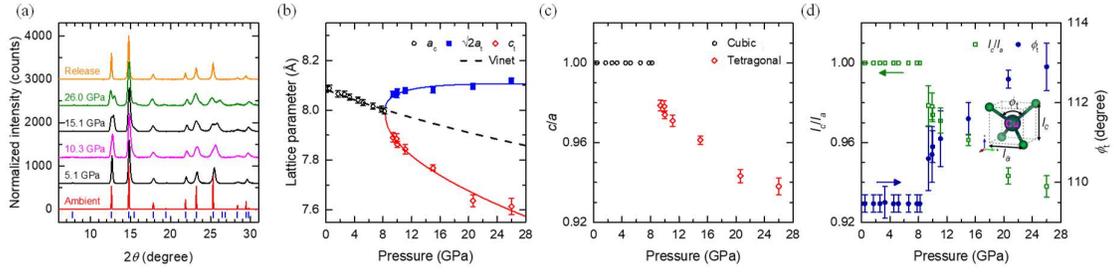

FIG. 3. (a) Pressure dependence of XRD pattern. Peak splitting of Bragg peaks emerges under hydrostatic pressure, and the split merges back as the pressure is released. (b) Pressure dependence of the lattice parameters. The cubic phase transforms into a tetragonal phase above $P_c=8(1)$ GPa. The lattice parameter of the cubic phase ($a_c$) was fitted to the Vinet equation of state (the dashed line), and those of the tetragonal phase ($a_t$ and $c_t$) were fitted to a mean-field function proportional to $(P/P_c)^\delta$ (the solid line). (c) Pressure dependence of a ratio of lattice parameters $c/a$. Above $P_c$, it drops down from unity. (d) Pressure dependence of the O-Cu-O angle $\phi_t$ of the CuO$_4$ tetrahedron and the ratio of the height $l_c$ to the base $l_a$ of the tetrahedron.

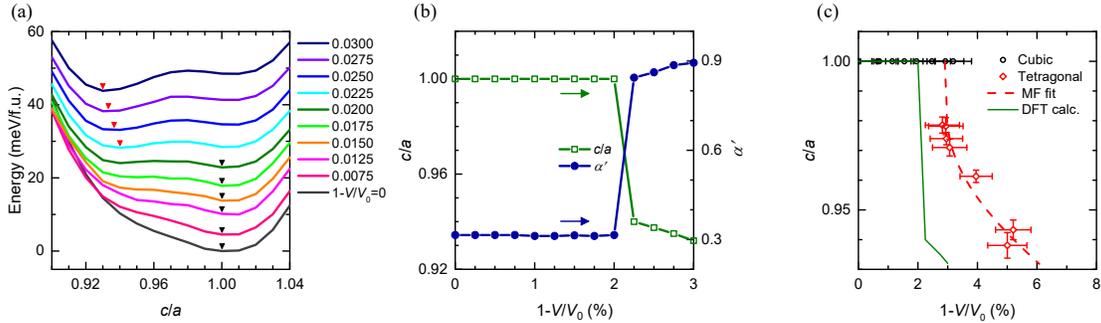

FIG. 4. (a) Calculated total energy depending on the change of the unit cell volume. The arrow denotes where the minima of the total energy locate. The global minimum shifts its position from $c/a=1$ to $c/a<1$ above the critical volume change of $1-(V_c/V_0)=0.02$. (b) Calculated volume change dependence of $\alpha'$ in Eq. (1) and $c/a$. The cubic's structural transition to the tetragonal lattice at $V_c$ accompanies a transition in the ground state from the $J_{eff}=1/2$ to $S=1/2$ states. (c) Comparison of $c/a$ ratio depending on the volume change between the experimental data points (symbols), two theoretical lines: one is that from the mean-field function (the dashed line) and another that from the DFT calculations in Fig. 3b) (the solid line).